\documentstyle[eqsecnum,aps,multicol,psfig,english]{revtex}
\include{header}
\begin{document}
\title{Energy Dissipation and Trapping of Particles Moving on a Rough Surface}
\author{C. Henrique$^1$, M.A. Aguirre$^2$, A. Calvo$^2$, 
I. Ippolito$^1$, S. Dippel$^3$, G. G. Batrouni$^4$ and D. Bideau$^1$}
\address{$^1$ Groupe Mati\`ere Condens\'ee et Mat\'eriaux (UMR 6626 au CNRS), 
Universit\'e de Rennes I, B\^at. 11B, Campus de Beaulieu, 35042 Rennes 
Cedex, France.}
\address{$^2$ Facultad de Ingenier\'ia. Paseo Col\'on 850, Buenos
  Aires, Argentina.}
\address{$^3$ H\"ochstleistungsrechenzentrum, Forschungszentrum J\"ulich, 
52425 J\"ulich, Germany}
\address{$^4$ Institut Non-Lin\'eaire de Nice,
Universit\'e de Nice-Sophia Antipolis, 
1361 route des Lucioles,
06560 Valbonne, France}

\maketitle

\begin{abstract}
  We report an experimental, numerical and theoretical study of the
  motion of a ball on a rough inclined surface. The control parameters
  are $D$, the diameter of the ball, $\theta$, the inclination angle
  of the rough surface and $E_{ki}$, the initial kinetic energy.  When
  the angle of inclination is larger than some critical value,
  $\theta>\theta_{T}$, the ball moves at a constant average velocity
  which is independent of the initial conditions. For an angle $\theta
  < \theta_{T}$, the balls are trapped after moving a certain
  distance. The dependence of the travelled distances on $E_{ki}$, $D$
  and $\theta$. is analysed. The existence of two kinds of mechanisms
  of dissipation is thus brought to light.  We find that for high
  initial velocities the friction force is constant. As the velocity
  decreases below a certain threshold the friction becomes viscous.
\end{abstract}

%\begin{multicols}{2} 
\section{\protect Introduction}
\label{sec:1}

Essentially due to the non linear and dissipative character of their
interactions, flowing grains exhibit a range of complex and
fascinating behaviour such as density waves~\cite{lee},
avalanches~\cite{bak}, arching~\cite{duran} and
segregation~\cite{guyon}. The energy exchange among the grains and
between grains and walls are particularly important, but not well
understood. The solid friction force itself is still an open
problem~\cite{heslot} dating back to the work of Coulomb. In addition,
the collective effects of this solid friction in a quasistatic
deformation of a granular solid are not well
understood~\cite{farhang}. To take into account the energy loss during
collisions, we use only a very empirical parameter -the restitution
coefficient- even if experiments~\cite{louge} have shown that this
coefficient is more complex than expected.

We have performed several experiments~\cite{Fx} on the motion of balls
on rough inclined planes. One of our most important results is that in
a regime characterized by an average constant velocity $V$, the
macroscopic friction force is "viscous", i.e. proportional to $V$.  In
the $D-\theta$ parameter space, where $D$ is the diameter of the
moving ball and $\theta$ the inclination angle, the constant velocity
regime, which we call regime $B$, is sandwitched between a
decelerated/trapping regime $A$ (for small $D$ and/or $\theta$) and
a regime C in which the ball jumps on the plane, and eventually can
reach a chaotic motion~\cite{alexa}. In this paper, we focus on how
the ball losses its energy in the decelerated/trapping regime, A.

After a short description of the experimental procedure, we present
results of the stopping distance of balls reached with low initial
kinetic energy, as well as results from an approximate two dimensional
stochastic model.  We discuss energy dissipation in
section~\ref{sec:3}, and in the last section we propose a
phenomenological model and compare it with numerical simulation.

\section{\protect Experimental Procedure}
\label{sec:2}

The experimental setup used in this work has been described in detail
elsewhere~\cite{ale,chris}. A rough plane is constructed by sticking
particles (see below) on an adhesive surface. This surface is placed
on a thick glass plane supported by a metal frame to prevent warping
and a jack that allows the accurate adjustment and measurement of the
angle of inclination, $\theta$, which is one of the control parameters
in our experiment. The rough surface is made of a monolayer of sifted
rolled sand grains with mean radius, $r$, $0.2 {\rm mm} \le r \le 0.25
{\rm mm}$. These grains are glued to the adhesive surface in such a
way as to obtain a homogeneous geometrically disordered rough surface
with a surface fraction compacity close to 0.8.  We used this surface
to study the motion of steel spheres of diameter $1.6 {\rm mm} \le D
\le 10.3 {\rm mm}$.

In order to control the initial kinetic energy supplied to the ball
(i.e. the initial speed with which the ball hits the rough surface) a
thin smooth plastic sheet is placed on the rough surface in the region
where the ball is launched. A set of parallel and equally spaced lines
is drawn on the sheet perpendicular to the direction of maximum slope
of the plane. The first line is drawn at a distance equal to the
smallest rolling ball radius, from the edge of the sheet nearest to
the rough surface.  To study the behavior of rolling balls with very
low initial kinetic energy, $E_{ki} \approx 0$, we release them from
this first line. To increase the initial $E_{ki}$, we simply release
the ball from one of the other lines traced higher up on the plane. It
is then trivial to see that $E_{ki}\propto X_i$, where $X_i$ is the
distance travelled over the smooth surface.

The experimental procedure is as follows. The inclination angle of the
plane is fixed at a desired value, and a straight barrier is placed at
the location which gives the desired initial kinetic energy,$E_{ki}$.
Twenty balls of a given radius, $R$, are positioned on the barrier on
the uphill side and not touching each other. Removing the barrier, the
balls are accelerated as they move the same distance, $X_i$, on the
smooth surface, arriving at the rough surface with the same initial
kinetic energy, $E_{ki}$.  Since we have set the inclination angle,
$\theta$, in the range corresponding to the pinning regime, $A$, the
released balls will come to a stop on the rough plane at various
distances $L_i$. The susbscript $i$ denotes a particular pinned ball.
This procedure is repeated $35$ times with the same experimental
settings to collect statistics. We then repeat the experiment for
different values of $E_{ki}$, $R$ and $\theta$, always keeping
$\theta$ in the pinning regime. The uncertainty in length measurements
is of the order of $\pm 1$ mm and the number of balls for each set of
parameters values investigated is large enough to have good
statistics.

\section{\protect Stopping distance of balls
with low initial kinetic energy} \label{sec:3}

%\subsection{\protect Experimental results} \label{subsec:3-1}

With the method described above, we measured the distance, $L_i$,
travelled on the rough surface by each ball and for a wide range of
control parameter values ($1.6 {\rm mm} \le D \le 10.6 {\rm mm}$ and
$1.5^{\circ} \le \theta \le 12.5^{\circ}$). In all cases $X_{i}$ is
taken as small as possible, around 2 {\rm mm}. In Figs.~\ref{expo} (a,
b) typical $L_i$ distributions are shown for $D = 2 {\rm mm}$ and
$\theta = 3^{\circ}$ and $5^{\circ}$ respectively. It can be seen that
these distributions peak for a travelled distance, $L_i$, of the order
of the diameter of the moving ball ($ 2{\rm mm}$). The distributions
decay for larger values of the travelled distance, the decay being
faster for the smaller $\theta$.

We found that the decay of these distributions for the larger values of
$L_i$ is well fitted by an exponential

\begin{equation}
\label{eqn:3-1}
N = N_0 e^{-\alpha L_i}.
\end{equation}

The solid lines in Figs.~\ref{expo}(a) and (b) are fits obtained with
$\alpha = 2.9 {\rm cm^{-1}}$ and $0.5 {\rm cm^{-1}}$ respectively.
Similar exponential distributions were found for the stopping distance
by Riguidel {\it et al},~\cite{Fx}, but working under experimental
conditions corresponding to regime B, where the moving ball reaches a
steady, albeit fluctuating, velocity.  In this regime, the balls are
occasionally stopped by large holes on the surface. Their results were
well fitted with $\alpha$ varying with $\theta$ as:

\begin{equation}
\label{pirulo1}
\alpha=exp\left(-a D^3 \sin^2\theta\right),
\end{equation}
where $a$ is a constant.

In order to complete and verify this law, we have performed a
systematic study of the variations of $\alpha$ with $\theta$.  In
Fig.~\ref{alpha} we show $\alpha$ as a function of $\theta$ for $D = 2
{\rm mm}$. Similar results were found for all the $D$ values studied.
Two different dependences are observed: A linear variation with large
negative slope for inclination angles lower than $4^\circ$ followed by
a crossover to a different smoothly decreasing curve tending to zero
for angles larger than $5^\circ$.  Good agreement was found using
eqn.~\ref{pirulo1} for this last part of the curve, as shown in the
figure (solid line). However, it is clear that eqn.~\ref{pirulo1} is
not valid for $\theta$ values smaller than $4^\circ$. This clearly
demonstrates the existence of a change in the trapping mechanism of
the rolling ball when the inclination angle goes from a value lower
than $4^\circ$ to a larger one. To find the angle at which the first
trapping mechanism disappears, we look for the intercept of the
straight line in Fig.~\ref{alpha} with the x-axis. We define this
transition angle, $\theta_T$, $\alpha=0$ as the angle for which the
trapping mechanism of regime A has disappeared.  In Fig.~\ref{alpha},
$\theta_{T} = 4.6^{\circ}$.  This value is very close to that found
previously~\cite{ale} for the angle $\theta_{AB}$ at which the
transition between the decelerated regime (regime A) and the mean
constant velocity one (regime B) occurs (for the same $D$).  In the
same way, we can find the angles $\theta_{T}$ corresponding to $\alpha
=0$, for all studied $D$ values. These $\theta_{T}$ angles can be
compared with the $\theta_{AB}$ values previously reported~\cite{ale}.
Very good agreement is obtained for large $D$ values, and slight
differences for smaller ones.

We can therefore conclude that there are two trapping mechanisms: One
due to large wells (disorder) in the surface, and the other one to
{\it dissipation} (controlled to a large degree by the relative
smoothness seen by the rolling ball). The measurement of the
dependence of $\alpha$ on $\theta$ described above, provides a more
physical and precise criterion for determining experimentally the
transition line between these two regimes.

The fact that in Fig.~\ref{alpha} the transition between the two
observed behaviours according to $\theta$ values occurs smoothly
indicates that near the transition, {\it both} trapping
mechanisms are important.

In fact, this experimental distribution of the stopping distances and
their exponential fits was predicted using a simple two
dimensional\footnote{The experiments are, of course, three
  dimensional: The two directions defining the rough plane, and the
  direction perpendicular to it.} stochastic model. The above
experimental verification came later. In this model one starts with
Newton's equations of motion for the moving balls which are then
simplified by making some approximations motivated by the geometry of
the collisions and the properties of the dissipation.  The resulting
model describes well the statistical properties (i.e. averages and
distributions) of physical quantities of the moving balls. For details
of the model see reference ~\cite{george}. In Fig.~\ref{alphaggb} we
show the $\alpha$ versus $\theta$ plot obtained from the stochastic
model. We see that it has the same form and functional dependence as
that found in the experiments. The actual values of $\alpha$ and
$\theta$ given by the model differ from the experimental ones. The
model is too simple to give accurate quantitative predictions, for
example rotation is ignored. However, the excellent qualitative
agreement between the predictions and the experimental results means
that the role of geometry, which is emphasized in the model is indeed
crucial.

%\subsubsection{\protect Distributions medians variations}
%\label{subsubsec:3-1-2}

We define the median, $L^*$, of the stopping distance distribution as
the length for which $50 \%$ of the balls are trapped on the rough
surface. Since we found exponential distributions in all cases we have
studied, we can write $L^*$ as:

\begin{equation}
\label{pirulo2}
L^*=\frac {\vert \ln(0.5)\vert} {\alpha}.
\end{equation}

As seen in Fig.~\ref{alpha} $\alpha= -a \sin(\theta)+b$, and from the
definition of $\theta_T$, $\alpha(\theta_{T})=0$, we see that
$\alpha_{\theta \to \theta_{T}}= a (\theta -\theta_{T})
\cos(\theta_{T})$, and therefore $L^*_{\theta \to \theta_{T}}\propto
\frac {1} {\vert \theta -\theta_{T} \vert}$.  Thus, $L^*$ diverges for
${\theta \to \theta_{T}}$ in a way reminiscent of a phase transition
with a critical exponent equal to unity.

In order to verify this behaviour we measure $L^*$ for the
distributions corresponding to different pairs of control parameter
values ($\theta, D$). In Fig.~\ref{L* le joli dessin} the dependence
of $L^*$ on $\theta$ is shown for three $D$ values. For $D=10.3 {\rm
  mm}$ the transition between regimes A and B occurs at
$\theta=2.5{^\circ}$ and for $D= 3 {\rm mm}$ at $\theta=4.4^{\circ}$.
In both cases the very rapid increase of $L^*$ as $\theta \to
\theta_T$ from below, demonstrates an approximate divergence
consistent with the above discussion. Let us recall, that as
$\theta\to\theta_T$ we have {\it two} competing pinning mechanisms and
therefore no true divergence. This ``divergence'' was based on the
idealization that as we approach $\theta_T$ from below, pinning in
regime A completely disappears. For the smallest $D$ value shown in
the figure ($D= 1.6 {\rm mm}$), the agreement between the proposed
functional dependence of $L^*$ on $\theta$ and the experimental
results is also clear, even if the A-B transition is not yet reached.

\section{\protect Energy dissipation in Regime A}
\label{sec:4}

In this section we discuss energy dissipation in regime A, i.e. small
angles of inclination, $\theta$, where the balls always come to rest.
With this goal in mind, balls were released from the various lines
marked on the plastic sheet, as described in section~\ref{sec:2}. In
this way, controlled initial kinetic energies were supplied to the
balls, and their stopping distances were studied.

We will first show that the transition A-B is not affected by the
initial kinetic energy. This is in agreement with previous
characterization of regime B as that interval of inclination angles
for which the balls reach an average steady state velocity {\it
independent} of the initial velocity (or kinetic
energy).\cite{george,sabine1,sabine2} We then analyze, for a very
small inclination angle ($\theta=2^\circ$), the stopping distance
distributions for different initial kinetic energies and several $D$
values. After that, we investigate the dependence of the mean distance
travelled by the ball with its initial velocity.  Finally, we propose
a phenomenological interpretation of the experimental results
obtained.

\subsection{\protect Experimental results}
\label{subsec:4-1}

To study the influence of the initial velocity ($V_{i}$) on the
transition between regimes A and B, 100 balls of diameter $ D = 3 {\rm
  mm}$ were released from $X_i = 0.3$ cm ($V_{i}\approx 5$~cm/s) and
then from $X_i = 20$ cm ($V_{i}\approx$~45~cm/s). The stopping
distance of each ball was measured, and $L^*$ calculated. This
procedure was repeated for ten different inclination angles of the
rough surface, $2.7^\circ \le \theta \le 6^\circ$, in order to insure
the change of the dynamical regime.

We have shown before that, with its rapid increase (``divergence''),
$L^*$ itself characterizes the transition. In Fig.~\ref{dosvelo} we
show this divergence of $L^*$ as a function of the inclination angle
for the two values of the initial velocity, $V_i=5,45$ cm/s. It is
clear that the A-B transition occurs at the same angle for both curves
independently of $V_{i}$. This is so even though the initial velocity
has a great influence on the travelling distance in regime A.

We now discuss the evolution of the distribution of stopping distances
as a function of the initial kinetic energy of the balls.  The
experimental procedure is as follows: The control parameters ($\theta$
and $D$) were fixed at values corresponding to regime A and a large
enough number of balls were released at different $X_i$.  The distance
travelled by every ball before being trapped was measured.  We show
the distributions for the stopping distances in Fig.~\ref{fantasmitas}
for $D =10.3 {\rm mm}$, $\theta= 2^\circ$ and four values of
$V_i$. The distributions are essentially the same, except that the
location of their centers moves towards increasing values of the
travelled distance as the initial velocity is increased. The
dispersion is very small.

We observe experimentally that those balls which begin their motion
with larger initial velocity travel a longer distance before getting
trapped. Moreover, balls released from the same $X_i$ (i.e. the same
initial velocity) move through almost the same distance on the rough
surface before getting trapped over a distance of a few centimeters:
The dispersion of the distribution of stopping distances is very
small. This leads us to postulate that the ball does not get trapped
unless its velocity is under some threshold value. To arrive at this
value, it must first travel a certain distance on the rough surface to
dissipate enough energy. Using a video camera and image processing, we
were able to evaluate this threshold value.  We launch several balls
($D=6 {\rm mm}$, $\theta =2^\circ$) with different initial kinetic
energies ($ 0.5 {\rm cm} \le X_i \le 18.5 {\rm cm}$) and for each
ball, the velocity is measured every acquired frame (15 frames by
second). Balls travel a distance of about 15~cm before getting
trapped. Our measurements indicate that balls do not get trapped while
their speed exceeds 3~cm/s. Note that the velocity below which the
ball can be trapped is independant of the initial
velocity\footnote{The same experiment was performed for $D=6 {\rm
    mm}$, and $\theta = 2.85^\circ$ the threshold obtained in this
  case is 6.5~cm/s.}. Since the distributions of stopping distances are
rather narrow and symmetric, the average value $\tilde{L}$ is a good
variable to use to characterize the energy loss. We show in
Fig.~\ref{rectas} a plot of $\tilde{L}$ as a function of $X_i$, for
$\theta= 2^\circ$ and $D=10.3,{\rm}$ and $4.7 {\rm mm}$.

The first thing to notice is that $\tilde{L}$ increases
with the initial kinetic energy ($\propto X_{i}$) of the balls. We
also see that for $D = 10.3 {\rm mm}$ and $X_i >$ 10~cm, this
dependence is linear. On the other hand, for $0 {\rm cm}\le X_{i} \le
10 {\rm cm}$, $\tilde{L}$ increases with $X_i$ following another
law. Note that for $X_{i} = 0$, $\tilde{L}$ must be 0. The same
behaviour is seen for $D= 4.7 {\rm mm}$ but, in this case, the
crossover between the two dependences is around $X_{i} = 2$~cm.

The presence of two different behaviours for the stopping distance as
a function of the initial kinetic energy suggests a change of the
nature of the friction force between the two regimes.  At the moment
we cannot characterize these forces any further.  However, in previous
work~\cite{Fx,george,ale,chris}, which concentrated on regime B,
friction mechanisms were extensively studied experimentally,
numerically and theoretically. Only three types of friction force are
possible: $F = K_1$, $F = K_2 V$ and $F = K_3 V^2$.  In the following
section we use these experimental results to propose a
phenomenological model with few parameters which reproduce the
experimental results.

\subsection{\protect Physical model} 
\label{subsec:4-2} 

Our objective is to describe the dynamics of a ball moving down a
rough slightly inclined surface. Taking into account the experimental
results just described, we can assume that two different types of
friction forces exist. Which type of force enters into play depends,
among other things, on the initial velocity of the ball. It is clear
that when moving down the plane, a ball loses its initial energy by
collisions and friction with the surface grains and finally is
trapped. However, not much is known about the way in which this
occurs. In particular, nothing is known about the average velocity as
a function of the distance travelled. So, the main assumptions of our
model are: (a) it is possible to define such a velocity function,
$V(D)$, for all balls, and (b) to get trapped, $V$ must be smaller
than a certain threshold value $V_{min}$. In other words, we are
assuming that to decrease its velocity from a value $V_0$ to a smaller
one $V_1$, a ball has to travel the same distance $\lambda_{01}$
independent of its initial velocity.

As for $V(D)$ we know that: $V(0) = V_i$ and we assume that
$V(L)~\leq~V_{min}$ where $V_i$ is the initial velocity of the ball
and $L$ is the corresponding trapping distance. Clearly, since the
physical quantities of interest are fluctuating quantities given by
some distribution, the quantities entering in the model are
statistical averages.

With the above considerations, we can write:

\begin{equation}
\label{cagada}
 \widetilde{L_0}=\widetilde{L_1} + \widetilde{\lambda_{01}},
\end{equation}
where $\widetilde{L_0}$ and $\widetilde{L_1}$ are the average stopping
distances of balls released with initial velocities $V_0$ and $V_1 <
V_0$, respectively, and $\widetilde{\lambda_{01}}$ is the mean
distance travelled by the first set of balls as their velocity
decreases from $V_0$ to $V_1$. Differentiating this equation with
respect to $V^2$, we obtain:

\begin{equation}
\label{dD}
 \frac {\partial \widetilde{\lambda(V)}} {\partial V^2}= - \frac {\partial
 \widetilde{L(V)}} {\partial V^2}
\end{equation}

The right hand side of eqn.~\ref{dD} may be evaluated from the
experimental results presented in the previous section
(Fig.~\ref{rectas}). As we have already mentioned, for large enough
initial kinetic energies, the dependence of $\tilde{L}$ on $V^2$ is
linear. Therefore, eqn.~\ref{dD} leads to: ${\partial V^2} / {\partial
  \widetilde{\lambda(V)}} = {\rm constant}$. This is the energy
gradient and in consequence must be equal to the force. We thus find
that, at large enough velocities, a ball moving down a rough surface
suffers a constant friction force, $F=mK$, and that the experimental
determination of $\widetilde{L(E_{ki})}$ provides a way to evaluate
it.

On the other hand, for smaller initial kinetic energies we have found
a different relation between $\tilde{L}$ and $V^2$ which, in fact,
implies another mechanism for energy dissipation. In this case, we
assume a viscous type friction force (i.e. proportional to the
velocity). For each pair of values ($\theta,D$), we define a
velocity, $V_l$, for which the crossover between the two types of
frictional forces occurs.

In other words, we propose that a ball of diameter $D$, moving down a
rough surface made of grains of mean radius $r$, and inclined an angle
$\theta$, suffers a constant dissipation force ($F=mK$), while its
velocity $V$ is larger than $V_{l}(\theta,D)$; a viscous type
dissipation force ($F=maV$) when its velocity is between
$V_{l}(\theta,D)$ and $V_{min}(\theta,D)$, and that it gets trapped on
the rough surface only if $V \leq V_{min}(\theta,D)$.

To verify these assumptions, we calculate the distances travelled
given by the above model and compare them to the experimental ones.
If a ball is released with a velocity $V > V_l$, the distance it
travels according to this model is easily calculated to be

\begin{equation}
\label{Xb}
\widetilde{L(V)}= \frac {1}{2K} \left(V^2 -V_l^2 \right)
   + \frac {V_l} {a}\left(1-\frac{V_{min}} {V_l} \right).
\end{equation}
On the other hand, if the release velocity, $V$, is less than $V_l$,
the distance travelled is

\begin{equation}
\label{Xs}
\widetilde{L(V)}= \frac {V} {a}\left(1-\frac{V_{min}} {V} \right).
\end{equation}

We arrive then, at two equations with four parameters: $a, K$ (which
characterize the friction forces), $V_{min}$ and $V_l$.

Using eqn.~\ref{dD}, $K(D)$ was calculated by fitting to the linear
part of the $\tilde{L}$ versus $V_i^2$ curves, as explained just after
Eqn.~\ref{dD}. Since the slopes of the linear regions in
Fig.~\ref{rectas} depend on $D$, $K$ is then a function of $D$.  To
determine $a$ and $V_{min}$ we fit Eqn.~\ref{Xs} to the curved regions
(small $X_i$) of Fig.~\ref{rectas}. However, in this region, we have
enough data for such a fit only for $D= 10.3 {\rm mm}$, which is what
we use. We note that for small $X_i$ (small initial velocity) the
stopping distance appears to be roughly the same for all $D$ values we
studied. We, therefore, take the values determined for $D= 10.3 {\rm
mm}$ as constant for all $D$.

Finally, the values of $V_l$ for the different $D$ values were
evaluated as follows. The constant (i.e. independent of $V$) term of
eqn.~\ref{Xb},

\begin{equation}
\label{j}
j = -\frac {V_l^2} {2K} + \frac {V_l} {a}\left(1-\frac{V_{min}}
{V_l} \right),
\end{equation}

can be obtained directly, for each $D$, from the intersections of the
fitted straight lines of $\tilde{L}$ versus $V_i^2$ with the vertical
axis. The solution of eqn.~\ref{j} gives two values of $V_l(D)$, which
are the two intersection points of equation~\ref{Xb} and \ref{Xs}. The
smaller one is taken as the physical one corresponding to a passage
from a more dissipative way to a less one.  Fig.~\ref{curvas} shows
the result of this analysis for all the studied values of $D$. Here we
plot the experimental average stopping distance, $\tilde{L}$, as a
function of the initial {\it velocity} (not $X_i\propto V^2$). The
solid lines show the calculated values using equations~\ref{Xs} and
~\ref{Xb}. It can be seen that the agreement is very good. Note that
since $a$ and $V_{min}$ have been taken as constants for all $D$, we
only have two parameters left for adjustment to fit the data for the
various $D$ values. So, the very good agreement with experiments is
obtained by tuning only two parameters.

\begin{table}[htbp]
\begin{center}
\begin{tabular}{|l|*{6}{c|}}
%\hline
$D ({\rm mm})$ & 2 & 3 & 4.7 & 6.3 & 7.1 & 10.3 \\
\hline
$K ({\rm m.s^{-2}})$ & 4.9 & 2.6 & 1.2 & 0.8 & 0.7 & 0.5 \\
\hline
$V_l ({\rm cm.s^{-1}})$ & 1.2 & 2.8 & 7.3 & 9.1 & 10.8 & 20.7 \\
%\hline
\end{tabular}
\end{center}
\caption{ Values of the different parameters, for $\theta = 2^\circ$}
\label{tab:1}
\end{table}

The values of the parameters are given in table~\ref{tab:1}.  The
value obtained for $V_{min}$, for the largest $D$, is consistent with
zero: it is not possible to give a $\tilde{L}$ value acceptable when
it is smaller than $D$, due to the measurements precision and the
width of the distributions. We have obtained $a=0.03 s^{-1}$ for the
prefactor in the viscous force while $K$ and $V_l$ values range
between $500 cm.s^{-2}$ and $50 cm.s^{-2}$ and 1~cm/s and 21~cm/s,
respectively, for values of $D$ increasing from 4 to 20.6. It can be
seen that the maximum value of the viscous deceleration, corresponding
to the largest $V_l$, is around $0.6 cm.s^{-2}$.

This means that a ball with $D= 10.3 {\rm mm}$ which begins its motion
over the rough surface with, for example, a velocity of 50~cm/s,
``feels'' a constant deceleration of $50 cm.s^{-2}$ until it reaches a
velocity of 21~cm/s. At that moment, the force {\it abruptly} becomes
viscous and the deceleration decreases to $0.6 cm.s^{-2}$. The
velocity then continues to decrease and so does the friction force.
When the velocity reaches the minimum value, $V_{min}$, the ball gets
trapped.

This behaviour is easily motivated physically. When the ball is
launched with a high velocity, collisions with the bumps on the
surface cause it to undergo rather large bounces. The time of flight,
$dt$, of these (ballistic) bounces is determined primarily by
$V_{\perp}$ (the velocity normal to the surface), $dt\propto
V_{\perp}$, and therefore the frequency of collisions is proportional
to $V_{\perp}^{-1}$. With each collision the ball losses an amount of
energy proportional to $V_{\perp}$, due to the coefficient of
restitution. Therefore the energy lost per second, the product of
these two quantities, is constant, i.e. a constant friction force.
More elaborate calculations find similar results.~\cite{alexa} 

At smaller velocities, the bounces are not high, the ball probes the
geometry of the surface and the motion is a mixture of bouncing and
rolling. The time of flight is, therefore, more complicated. The
motion closely resembles that in the constant velocity regime thus
giving viscous friction. The difference between this motion and the
constant velocity regime, is that here the energy gained by the ball
moving down the slope cannot compensate for the energy lost. The ball
eventually gets pinned.

$V_{l}$ appears like a limit velocity at which the ball can
just ``fly-over'' the grains constituting the surface.

\section{\protect Numerical simulations}
\label{sec:5}

To investigate in more detail our conjectures concerning the motion of
the particle we performed numerical simulations of the system. The
motion of the particle was simulated using soft sphere molecular
dynamics (for details see \cite{dip97}). The sphere moves on a plane
configuration scanned in from one of those used by Riguidel in his
experiments \cite{Fx}. As material parameters we use a normal
coefficient of restitution $e_n=0.6$ and a coefficient of friction of
$\mu=0.13$. 

First we checked that we find the same global behaviour as in the
experiments. Figure~\ref{Lsim} shows the mean stopping distance, as
averaged over the launching of 60 balls per starting velocity as a
function of the initial kinetic energy. We find the same behaviour as
in experiments -- a linear region for higher starting velocities, and 
a not so well defined different shape of the curve for smaller
starting velocities.  

We now want to check whether the ball really covers the largest part
of $\tilde{L}$ in large jumps. Figure \ref{fldist} shows the distance
covered between jumps as a function of time for different starting
velocities. Obviously, the first part of the motion consists of very
wide jumps (covering a few particle diameters at high starting
velocities), but then drops very rapidly to much smaller distances.
The times between collisions suddenly exhibit the very regular
behaviour also observed in simulations of the steady state motion
\cite{dip97,dip96}: a number of small jumps, in the course of which
all normal velocity with respect to the ball on the plane is lost,
followed by a rolling over the rest of the ball. With the small angles
of inclination and velocities in this case, this rolling starts very
early on the ball, as is obvious from the long distances between
collisions corresponding to them. In Fig.~\ref{fldistint} the total
distance covered in the same runs as in Fig.~\ref{fldist} is shown.
Here, it becomes obvious that indeed, most of the $\tilde{L}$ is due
to the large jumps at the beginning of the motion, and we see a very
clearly defined crossover in the friction force. Only the curve
corresponding to Fig.~\ref{fldist}(a) does not exhibit this crossover.
The reason is that the initial jumps were already not much longer than
a particle diameter, since the initial velocity was quite low. Thus,
the discrepancy between the two regimes of the motion is not very
strong. Besides, the starting velocity corresponding to
Fig.~\ref{fldist}(a) is quite close to the lower limit of the linear
region in Fig.~\ref{Lsim}, i.e.~close to a different type of
behaviour.

 From Fig.~\ref{fldist} it is also obvious that the stopping {\em time}
is very similar in all cases, but that the onset of rolling appears a
bit later for higher starting velocities. As we assume in our physical
model in section IV.B., stopping only takes place after the ball
velocity has dropped below a certain value (in our case somewhere
around 7 cm/s), at which point the rolling starts. This can be seen in
Fig.~\ref{vel}, where we have plotted the evolution of the x-velocity
of the ball for the same cases as in Fig.~\ref{fldist}. The point
where the ball starts to move in bounces much smaller than a ball
radius (the stage prior to rolling) is marked with a small arrow
for each trajectory. 

It can also be seen from these curves that after a very rapid drop in
the velocity after the first few collisions with the plane, the
velocity seems to decrease in a linear fashion, though with a slope
that seems to depend slightly on the initial velocity. Then, when the
particle enters the phase of the motion consisting of a number of
bounces with each ball it passes (and eventually some rolling), the
friction force experienced by the particle drops to a much lower value
and seems to be independent of the initial velocity.

\section{\protect Conclusion}
\label{sec:6}
We have investigated by means of experiments and numerical simulations
the movement of a single ball on a rough inclined plane, at small
inclination angle.  This has given us a better understanding of the
energy dissipation at small angles. We have shown that a ball which
has a large enough initial velocity $V_0$ first bounces on the rough
surface and suffers a constant friction force. Clearly in this case
the ball cannot be trapped if its velocity is larger than the
crossover velocity $V_l$. When the velocity reaches $V_l$, the
friction force suddenly becommes viscous: the dynamics of the motion
is now similar to that observed in regime B.  The key for
understanding these two mechanisms of dissipation, i.e. friction
forces, is the difference in the nature of the collisions when the
velocity is above or below $V_l$ as explained above. We have also shown
that the geometry of the surface plays an important role in the trapping 
of the ball. In regime {\it A} the ball is first slowed down gradually and
when the velocity finally reaches a threshold value, $V_{min}$, (which
appears to be {\it independent} of initial conditions) the ball is
trapped. The trapping probability decreases linearly with the
inclination of the plane. For the transition angle $\theta _{T}$, this
trappping mechanism disappears and the ball crosses over into the
dynamic regime {\it B} where it moves on the plane with a constant
mean velocity and is subjected to a viscous friction
force~\cite{Fx,ale}. In this regime, the ball can still be trapped by
the occasional big defect on the surface but its trapping probability
decreases {\it exponentially} with the angle of inclination. The fact
that in both regimes {\it A} and {\it B} the friction force is viscous
just before the ball gets trapped emphasizes the important fact that
the difference between the two regimes {\it A} and {\it B} is not of
dynamic origin, but due to two different trapping mechanisms.

The ``divergence'' of the median, $L^*$, of the stopping distance
(defined as the length for which $50\%$ of the balls are trapped) at
$\theta_{T}$ is seen to indicate clearly the transition between the
$A$ and $B$ regimes.  The values of $L^*$ depend on the initial
velocity but the divergence always occurs at the same angle
$\theta_{T}$. This shows that the location of the transition is
independant of initial kinetic energy.  Numerical simulations gave us
``microscopic'' details of the motion (like the time between
collisions) which agreed with and confirmed the experimental
measurements and our explanations.
%\end{multicols}
%

%
%%%%%%%%%%%%%%%%%%%%%%%%%%%%%%%%%%%%%%%%
\begin{figure}[htbp]
%\centerline{\psfig{....,width=8cm}}
\caption{Histogram of the stopping distance for
  $D=2mm$ (a) $\theta=3^\circ 5$, (b) $\theta=5^\circ$, $X_{i} \approx
  2mm$}
\label{expo}
\end{figure}
\begin{figure}[htbp]
%\centerline{\psfig{....,width=8cm}}
\caption{Decay constant of the stopping distance distributions $\alpha$ 
as a function of the rough surface inclination angle $\theta$, for 
$D=2 {\rm mm}$.}
\label{alpha}
\end{figure}
\begin{figure}[htbp]
%\centerline{\psfig{....,width=8cm}}
\caption{Same as in figure~\ref{alpha} but obtained from the stochastic model,
[12]}
\label{alphaggb}
\end{figure}
\begin{figure}[htbp]
%\centerline{\psfig{....,width=8cm}}
\caption{Median distributions, $L^*$, as a function of $\theta$ 
for ($\diamond) D = 10.3 {\rm mm}, (\circ) D =3 {\rm mm}, 
(\triangleright) D= 1.6 {\rm mm}$}
\label{L* le joli dessin}
\end{figure}
\begin{figure}[htbp]
%\centerline{\psfig{....,width=8cm}}
\caption{Variations of $L^*$ with $\theta$, for $D = 3{\rm mm}$, 
and two initial velocities: $(\circ) V_i \approx 5$~cm/s and 
$(\bullet) V_i \approx 45$~cm/s.}
\label{dosvelo}
\end{figure}
\begin{figure}[htbp]
%\centerline{\psfig{....,width=8cm}}
\caption{Distributions of the stopping distance for $D = 10.3{\rm mm} $, 
$\theta=2^\circ$, and different initial square velocities 
$V_i^2 \propto X_i$, from left to right $X_i = 0.5, 4.5, 10.5, 
16.5{\rm cm} $}
\label{fantasmitas}
\end{figure}
\begin{figure}[htbp]
%\centerline{\psfig{....,width=8cm}}
\caption{Average stopping distance $\tilde{L}$ as a function of 
$V_i^2 \propto X_i$, for $\theta= 2^\circ$, $ (\triangle)D= 10.3
{\rm mm}$ and $(\circ) D= 4.7{\rm mm}$}
\label{rectas}
\end{figure}
\begin{figure}[htbp]
%\centerline{\psfig{....,width=8cm}}
\caption{Average stopping distance $\tilde{L}$ as a function of 
$V_i$ in arbitrary units, for $\theta= 2^\circ$, $(\triangle) 
D= 10.3{\rm mm}$, $(\circ) D =7.1{\rm mm}$, $(\triangleright) 
D= 6.3{\rm mm}$, $(\bullet) D= 4.7{\rm mm}$, $(\triangleleft) D = 3 
{\rm mm}$ and $(\diamond) D= 2 {\rm mm}$. In filled lines the 
calculated variation for the corresponding $D$ values.}
\label{curvas}
\end{figure}
\begin{figure}
\caption{Average stopping distances for $D=5$, $\theta=1^\circ$.}
\label{Lsim}
\end{figure}
\begin{figure}
\caption{Distance covered by the ball between sucessive collisions 
with the plane for three different starting velocities (a) $v_i=20 
{\rm cm/s}$, (b) $v_i=30 {\rm cm/s}$, $v_i=40 {\rm cm/s}$. The 
diameter of balls on the plane is 1~mm.}
\label{fldist}
\end{figure}
\begin{figure}
\caption{Total distance covered by the ball as a function of time for
  the same cases as in Fig.~\ref{fldist}. Full line corresponds 
to (a), dashed line to (b), dot-dashed line to (c).}
\label{fldistint}
\end{figure}
\begin{figure}
\caption{Velocity of the ball (same as in Fig.~\ref{fldist}). Full
  line corresponds to (a), dashed line to (b), dot-dashed line to
  (c).} 
\label{vel}
\end{figure}
\end{document}